# Observation of Three Mode Parametric Interactions in Long Optical Cavities

**C. Zhao, L. Ju, Y. Fan, S. Gras. B. J. J. Slagmolen[*], H. Miao, P. Barriga D.G. Blair,**

School of Physics, The University of Western Australia, Crawley, Western Australia, 6009 Australia

**D. J. Hosken, A. F. Brooks, P. J. Veitch, D. Mudge, J. Munch,**

Department of Physics, The University of Adelaide, Adelaide, South Australia, 5005 Australia

We report the first observation of three-mode opto-acoustic parametric interactions of the type predicted to cause parametric instabilities in an 80 m long, high optical power cavity that uses suspended sapphire mirrors. Resonant interaction occurs between two distinct optical modes and an acoustic mode of one mirror when the difference in frequency between the two optical cavity modes is close to the frequency of the acoustic mode. Experimental results validate the theory of parametric instability in high power optical cavities.

The principles of parametric interactions have been widely used in physics, including low noise microwave amplifiers, optical parametric amplifiers (OPAs) and optical spring interactions with mechanical resonators. In the case of opto-mechanical interactions a mechanical mode modulates the length of an optical cavity, thereby changing the resonance condition of the optical mode. Such interactions have been observed in [1, 2, 3] and have recently been used to "cool" thermally excited mechanical modes of small acoustic resonators through the associated time dependent radiation pressure forces acting on the resonator [4, 5, 6, 7, 8]. Such techniques are examples of two-mode parametric interactions in which the linewidth of the optical mode is sufficiently broad that the mechanical frequency occurs within the linewidth of a single mode. The phenomenon reported here is analogous to stimulated Brillouin scattering (SBS) [9, 10]. In this case an optical mode from a resonator excites an acoustic hypersonic wave in a material at optical wavelengths, from which a second optical mode in the same resonator is excited by scattering off the moving sound grating. Three mode scattering into symmetric cavity longitudinal modes has been studied in a LIGO interferometer [11]. To cause parametric instabilities the scattering must occur into cavity modes which are distributed asymmetrically (in frequency space) about the carrier

---

[*] Now at Centre for Gravitational Physics, The Australian National University, Canberra, 0200, Australia

frequency. In addition, the transverse structure of the optical mode must match an acoustic mode structure.

Here we present the first observation of such three-mode opto-acoustic parametric interactions. In this case a cavity longitudinal mode and a transverse mode interact with a low frequency acoustic mode of a mirror. The energy in the cavity longitudinal mode (carrier) is scattered into the transverse mode (sideband) by its interaction with an acoustic mode in one of the cavity mirrors. In high-power and low-loss systems, this interaction can lead to parametric instability.

A one-dimensional analysis of three-mode opto-acoustic parametric interactions in the context of advanced gravitational wave interferometers was described by Braginsky *et al.* [12, 13]. Their analysis was extended by Zhao et al [14] to include the 3D acoustic mode structure of the mirrors and the optical cavity mode shapes. Ju et al [15] went on to consider the effect of parametric scattering into multiple optical modes.

These analyses predict that many of the features that optimise the sensitivity of advanced gravitational wave detectors also increase the likelihood of the parametric excitation of its mirrors. In particular, the new generation of advanced interferometers will use large mirrors that have low acoustic losses to reduce the effects of radiation pressure fluctuations and Brownian noise, respectively. They use long optical cavities and extremely high stored power to reduce the effect of shot noise. The resulting high optical and acoustic mode densities plus high optical power, can lead to a risk of parametric excitation of the mirrors without careful design. It is therefore important that the above analyses be validated so that the practical importance of parametric instabilities can be assessed. Here we present experimental results for three mode parametric interactions. Measurements are in agreement with the theory [12] and consistent with specific predictions for this experiment [16]

Parametric interactions can be considered as simple scattering processes [17], as indicated in Fig. 1. In (a) a photon of frequency $\omega_0$ is scattered, creating a lower frequency (Stokes) photon of frequency $\omega_s$ and a phonon of frequency $\omega_m$, which increases the occupation number of the acoustic mode. In (b) a photon of frequency $\omega_0$ is scattered from a phonon creating a higher frequency (anti-Stokes) photon of frequency $\omega_a$, which requires that the acoustic mode is a source of phonons, thus reducing its occupation number. The scattering could create entangled pairs of phonons and photons [18].

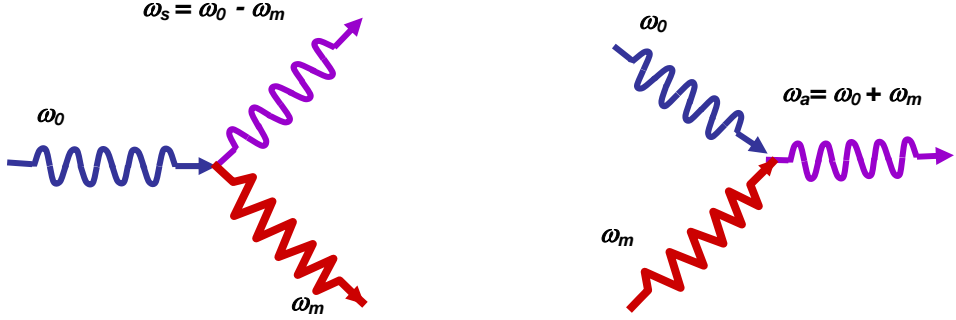

**Figure 1** Parametric scattering of a photon of frequency $\omega_0$ a) into a lower frequency Stokes photon, $\omega_s$, and a phonon of frequency $\omega_m$, and b) into a higher frequency anti-Stokes photon, $\omega_a$, which require destruction of a phonon.

This three mode interaction can only occur strongly if two conditions are met simultaneously. Firstly, the optical cavity must support eigenmodes that have a frequency difference approximately equal to the acoustic frequency: $|\omega_0 - \omega_{s(a)}| \approx \omega_m$. Secondly, the optical and acoustic modes must have a suitable spatial overlap.

The effect of the parametric scattering can be characterised by a dimensionless parametric gain, R, [13] given in Eq 1. In general, there may be multiple Stokes and anti-Stokes interactions for each acoustic mode. For simplicity, we include only one of each interaction type [13]:

$$R = \frac{4PQ_m}{McL\omega_m^2}\left[\frac{Q_s \Lambda_s}{1+(\Delta\omega_s/\delta_s)^2} - \frac{\omega_a}{\omega_s}\frac{Q_a \Lambda_a}{1+(\Delta\omega_a/\delta_a)^2}\right] \quad (1)$$

Here $Q_m$ is the quality factor of the acoustic mode; $M$ is the mass of the mechanical resonator; $P$ is the power stored in the TEM$_{00}$ mode; $L$ is the length of the cavity; $\Delta\omega_{s(a)} = |\omega_0 - \omega_{s(a)}| - \omega_m = \Delta_{s(a)} - \omega_m$; $\omega_{s(a)}$, $Q_{s(a)}$ and $\delta_{s(a)} = \omega_{s(a)}/2Q_{s(a)}$ are the frequencies, Q-factors and linewidths of the Stokes (anti-Stokes) optical modes; $\Delta_{s(a)}$ is the difference between the TEM$_{00}$ mode and the higher-order Stokes (anti-Stokes) mode; and the overlap factors $\Lambda_{s(a)}$ are given by [13]

$$\Lambda_{s(a)} = \frac{V(\int f_0(\bar{r}_\perp) f_{s(a)}(\bar{r}_\perp) u_z d\bar{r}_\perp)^2}{\int |f_0|^2 d\bar{r}_\perp \int |f_{s(a)}|^2 d\bar{r}_\perp \int |\bar{u}|^2 dV} \qquad (2)$$

where $f_0$ and $f_{s(a)}$ describe the optical field distribution over the mirror surface for the TEM$_{00}$ mode and higher-order modes respectively, $\bar{u}$ is the spatial displacement vector for the mechanical mode, and $u_z$ is the component of $\bar{u}$ normal to the mirror surface. The integrals $\int d\bar{r}_\perp$ and $\int dV$ correspond to integration over the mirror surface and the mirror volume $V$ respectively. The experiment described below investigated scattering processes between the TEM$_{00}$ mode and an anti-Stokes TEM$_{01}$ mode, so hereafter we omit the subscripts for $\Delta\omega$.

The gain R characterizes the effect of the parametric interactions on the acoustic mode. Positive R corresponds to interactions dominated by Stokes modes, while negative R corresponds to interactions dominated by anti-Stokes modes. For positive R < 1 the amplitude of the acoustic mode is increased by a factor 1/(1-R). If R > 1, the amplitude should increase exponentially with time until non-linear losses lead to saturation. For negative values of R, energy is extracted from the acoustic mode and if it is normally in thermal equilibrium, the effective mode temperature is cooled to a value of ~T/R, where T is the physical temperature of the acoustic resonator [6]. For R < 1 the interaction also causes changes in the relaxation time of the acoustic mode.

The 3-mode interaction could thus be observed by monitoring either the amplitude or relaxation time of the acoustic mode, or by measuring the amplitude of the anti-Stokes optical mode. In the experiment we report here, the maximum gain accessible was R~10$^{-2}$ and therefore we chose to measure the power scattered into an anti-Stokes TEM$_{01}$ mode of an optical cavity as the frequency difference $\Delta\omega$ between the TEM$_{00}$ and TEM$_{01}$ mode was tuned across the 160kHz acoustic resonance of a cavity mirror. The mirror was electrostatically excited using a comb capacitor [19] placed near the mirror bottom back surface.

The layout of the measurement system is shown in Fig. 2. The optical cavity consists of two sapphire mirrors suspended 80 m apart on simple wire-loop pendula in a vacuum system, yielding a free-spectral range of about 1.9 MHz. The physical and optical parameters of the cavity are listed in Table 1. The mirrors are aligned to produce a Fabry-Perot cavity by applying magnetic forces to small magnets glued to the back of the mirrors [20].

Table 1. Parameters of the cavity

|  | ITM | ETM | CP |
|---|---|---|---|
| Radius of Curvature (m) | Flat | 720 | flat |
| Materials | Sapphire | Sapphire | Fused silica |
| Diameter (mm) | 100 | 150 | 160 |
| Thickness (mm) | 46 | 80 | 17 |
| HR transmission (ppm) | 1840±100 | 20 |  |
| AR reflectivity (ppm) | 29±20 | 12±12 | 100 |
| Cavity internal power (kW) | 1.0 | | |
| Cavity length (m) | 77 | | |

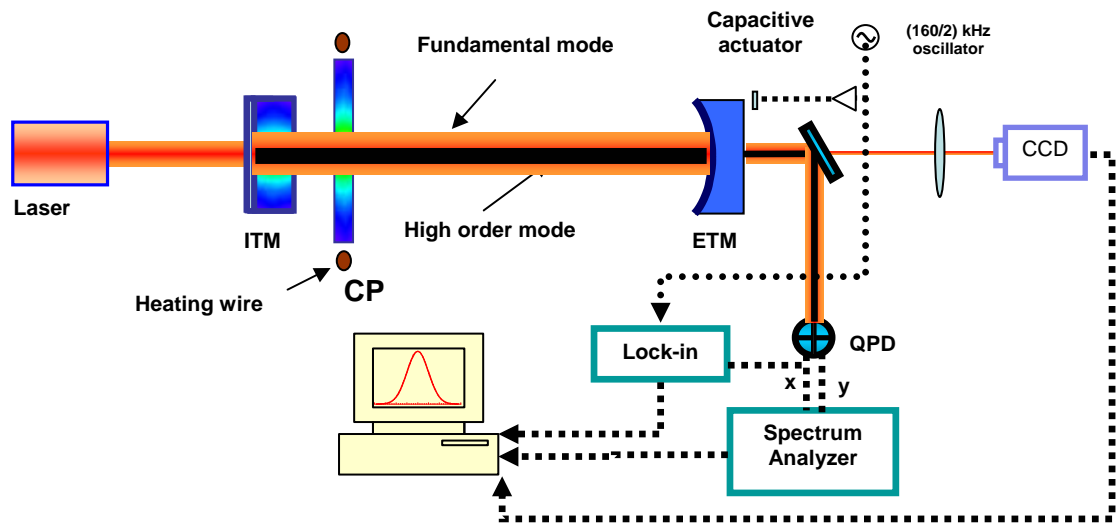

**Figure 2** Schematic diagram of the measurement system. The single-frequency 10W Nd:YAG laser was phase-locked to a $TEM_{00}$ mode of a 80-meter suspended-mirror cavity. The end test mirror (ETM) was resonantly excited at the acoustic resonance. The fused silica compensation plate (CP) was heated around its cylindrical surface, thereby inducing a negative lens, which tuned the frequency difference between the $TEM_{00}$ and $TEM_{01}$ modes. The cavity tuning and the anti-Stokes $TEM_{01}$ mode excitation were measured using the CCD and quadrant photodiode (QPD) at the back of ETM.

The single frequency Nd:YAG laser was phase-locked and mode-matched to the $TEM_{00}$ mode of the optical cavity, providing about 4W at the input test mirror (ITM). The $TEM_{00}$ finesse of the optical cavity is $1.3 \times 10^3$, limited mostly by the absorption losses in the substrate of the rear-surface ITM, giving an intra-cavity $TEM_{00}$ power of about 1 kW. The nominal cold-cavity beam radius at the end test mirror (ETM) is 9.2mm, which reduces by thermal lensing to 8.6mm at full power.

The acoustic mode shape and hot-cavity overlap factor were calculated using finite element modelling. The calculated contour map of the normal component at the ETM surface for the 158.11 kHz mode is shown in Fig. 3(a). The acoustic mode used for the measurement has a resonant frequency of 159.96 kHz, acceptably close to the calculated frequency, considering the magnets and suspension wire attached to the ETM and the accuracy of the material parameters [13], and a $Q \approx 7 \times 10^5$. The overlap of this mode is indicated in Fig. 3(b), in which a vertical cross-section of the product of the mode amplitude and the $TEM_{00}$ mode is compared to the $TEM_{01}$ mode amplitude. The overlap factor for this mode is 1.67, assuming that the optical mode is aligned with the geometric centre of the ETM.

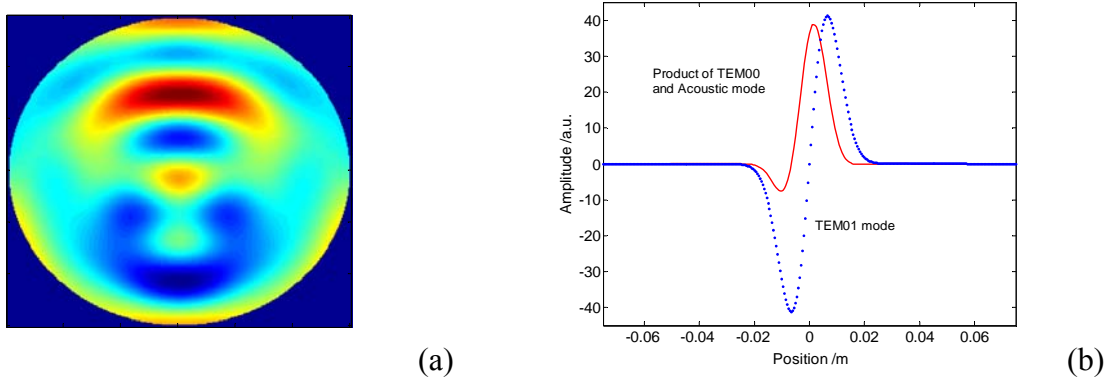

(a)                                               (b)

**Figure 3** (a) A contour map of the normal component at the ETM surface of the acoustic mode at 158.11 kHz; (b) a vertical cross-section of the product of the mode amplitude and the $TEM_{00}$ mode is compared to the $TEM_{01}$ mode amplitude.

Tuning of the frequency of the $TEM_{01}$ mode relative to that of the $TEM_{00}$ mode was accomplished using the intra-cavity low-absorption fused silica compensation-plate (CP), see Fig. 2. Heating the CP around its cylindrical surface creates a negative thermal lens, thereby changing the g-factor of the cavity, or equivalently, the effective curvature of the ITM. The frequency difference between the $TEM_{00}$ mode, $\omega_0$, and the anti-Stokes $TEM_{01}$, $\omega_a$, is given by

$$\Delta\omega = \omega_0 - \omega_a = \frac{c}{L}\left(\arccos\sqrt{\left(1-\frac{L}{R_1}\right)\left(1-\frac{L}{R_2}\right)}\right) \qquad (3)$$

where $R_1$ is the effective radius of curvature of the ITM and $R_2$ is the radius of curvature of the ETM. Heating the CP with about 15W over about 2 hours increased the g-factor from its nominal hot-cavity value of 0.87 to greater than 0.99.

The tuning of the cavity and the power in the $TEM_{01}$ mode were measured using the optical power leaking through the ETM. Some of this power was focussed by a lens to create an image of the ETM spot onto a CCD camera. The spot size was used to calculate the effective curvature of the ITM and thus the frequency difference between the $TEM_{00}$ and $TEM_{01}$ modes.

The power in the $TEM_{01}$ mode was determined by using a differential readout of the quadrant photodiode (QPD) to measure the heterodyne beat between the overlapping $TEM_{00}$ and $TEM_{01}$ modes. The differential readout discriminates against spurious signals due, for example, to direct acoustic modulation of the $TEM_{00}$ mode or electromagnetic pick-up. Note that any spurious signals should be largely independent of the thermal tuning of the CP.

The square of the QPD voltage component at the acoustic mode frequency, which is proportional to power in the $TEM_{01}$ mode, is plotted as a function of the frequency difference between the $TEM_{00}$ and $TEM_{01}$ modes, $\Delta\omega$, in Fig. 4. Fig. 4 also shows the parametric gain predicted by Eq 1, using the measured input power, calculated overlap parameter, and measured Q-factors including the $TEM_{01}$ mode linewidth that was independently measured to be 1.46 kHz. The noise in the measured data is due to fluctuations in the cavity alignment, which affects the circulating power, the mode overlap and the beam spot size measurement from which the mode spacing is inferred.

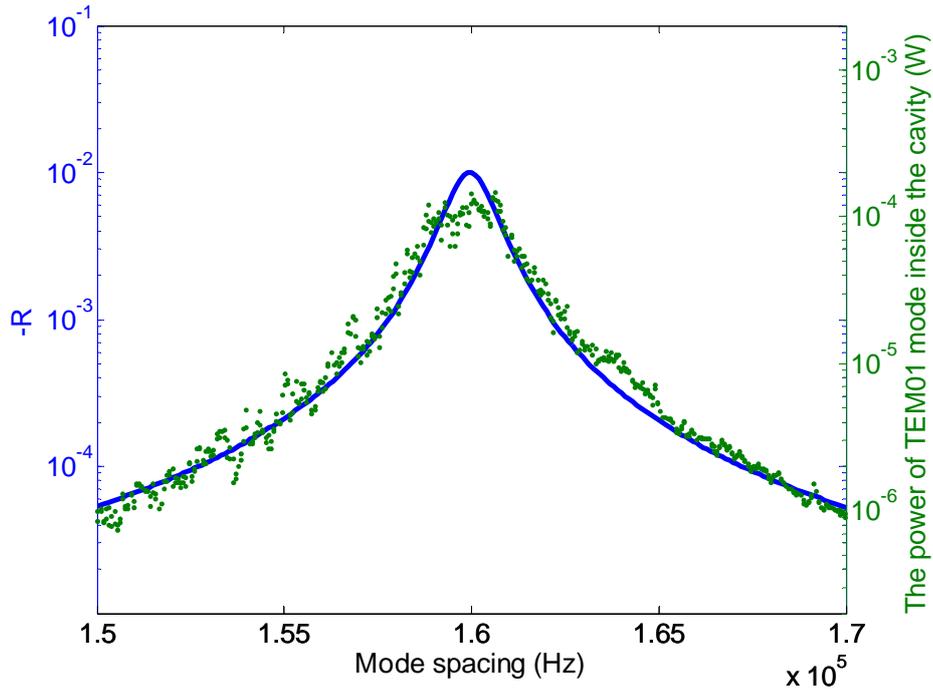

**Figure 4** Plot of the measured power of the TEM$_{01}$ mode (dots) as a function of the frequency difference between the TEM$_{00}$ and TEM$_{01}$ modes (mode spacing). The peak power occurs at the frequency difference corresponding to a cavity g-factor of 0.967. The solid line is the parametric gain predicted by Eq (1) using the calculated overlap factor, the independently measured linewidth of both optical modes, the input power, and the measured acoustic mode Q-factor. The related calibration between the two y-axes is estimated but due to uncertainty in the acoustic excitation amplitude, has been corrected by a multiplicative parameter.

We observed similar resonant parametric gain for another acoustic mode at ~84 kHz. For this mode and the 160 kHz mode, we confirmed the linearity of the readout by observing an accurate exponential decay of the mirror mode using cavity readout. For the 84 kHz mode we also used an independent stress birefringence readout [21] of the acoustic mode to verify that the high order mode signal is a true readout of the acoustic mode. Figure 5 shows the time trace for one of these measurements.

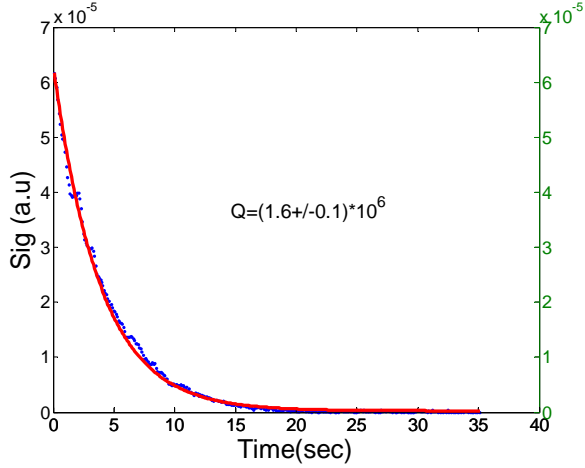

**Figure 5** The recorded time trace of the QPD output demodulated at the 84 kHz acoustic mode frequency (dots) after the electrostatic excitation was stopped at time zero. The solid curve shows the expected exponential decay for the 84kHz acoustic mode which had a Q-factor of $1.6\times10^6$.

The magnitude of R is proportional to a product of three quality factors since the circulating power P scales with the $TEM_{00}$ cavity loss. Our system has mirror Q-factors limited by glued-on magnets, and optical losses limited by absorption in the internal substrate of the ITM and Fresnel reflection at the compensation plate. Nevertheless, we are able to observe the three mode parametric interaction enhanced by the cavity resonance. The corresponding parametric gain as shown in Figure 4 is 0.01. This is insufficient to substantially change the acoustic mode relaxation time. However R ~100 could be achieved by using cavity finesses ~ $10^4$, acoustic mode Q's ~ $4 \times 10^7$ and a laser power of 20W. For a Stokes mode, this gain would represent a severe parametric instability.

In conclusion, we have observed, for the first time, a three-mode parametric interaction in a long optical cavity in which photons in a $TEM_{00}$ optical mode are scattered by resonant acoustic phonons into a $TEM_{01}$ mode. The system is analogous to SBS and also to an optical parametric amplifier OPA except that one optical mode is replaced with an acoustic mode and the non-linear interaction is via radiation pressure. We describe the system as an opto-acoustic parametric amplifier (OAPA). By direct comparison with the OPA, the OAPA could be a source of phonon-photon entanglement and could find applications in quantum information, teleportation, and quantum encryption [18]. Much effort has already gone into defining methods for suppressing parametric instability in advanced gravitational wave detectors. A combination of low noise damping rings [22] and stable power recycling cavity design [23] is likely to lead to stable solutions.

This investigation was conducted at the High Power Test Facility (HPTF), located at Gingin in Western Australia [24] which was developed in a collaboration between the Australian Consortium for Gravitational Astronomy and the US LIGO Laboratory. We would like to thank the International Advisory Committee of the Gingin Facility for their encouragement and advice, the centre of gravitational physics of Australian National University for useful discussions and for providing cavity locking electronics. This research was supported by the Australian Research Council and the Department of Education, Science and Training and by the U.S. National Science Foundation.